# Molecular Association at the Microscopic Level

Richard M. Neumann*
The Energy Institute
The Pennsylvania State University
University Park, Pennsylvania 16802
(3/12/05)

**Abstract**

The Helmholtz free-energy $W$ is calculated as a function of separation distance for two molecules in a fluid, A and B, whose mutual interaction is described by a spherically symmetric potential. For the equilibrium A + B $\Leftrightarrow$ AB occurring in a dilute solution or gas, $W$ is used to evaluate the association constant $K_{AB}$, which for ionic A and B is identical to the Bjerrum result. The criterion defining the bound species is not arbitrary; i.e., the cutoff separation distance in the configuration integral used to calculate $K_{AB}$ arises directly from the definition of $W$. For a one-component, dense fluid, $W$ permits the derivation of the phase-condensation temperature, which for a gas is the critical temperature and for a liquid the freezing temperature. For ionic A and B (e.g., $Na^+$ and $Cl^-$ in molten NaCl), an expression for the freezing temperature is obtained, which is similar to the expression for the melting temperature derived by Kosterlitz and Thouless for two-dimensional systems.

*Email: rmn4@psu.edu

## Introduction

There is considerable renewed interest in the calculation of the association (equilibrium) constant for the reaction A + B ⇔ AB, $K_{AB}$ = [AB]/[A][B], from first principles, primarily because of the many biological reactions where A and B are macromolecules. The bracketed quantities indicate concentrations. In one approach valid for A and B in the gas phase or in a liquid solvent, the expectation value of $K_{AB}$ is obtained directly from the configuration integral (CI) using a potential of mean force.[1-3] However, even in the simplest system where A and B are small molecules interacting with each other through a spherically symmetric potential, a conceptual difficulty appears, namely the definition of what constitutes the bound-species AB[1-5]. The evaluation of the CI requires a means of determining that region of phase space which is associated with the existence of AB in order to estimate the cutoff-separation distance, or upper limit of integration, in the CI. In his consideration of the association of ionic species, Bjerrum[4] chose the cutoff (also called the association distance) to be the distance where the work of separation against the coulomb force is twice the thermal energy; beyond this separation the ions are regarded as free. This choice remains controversial because the factor of two is arbitrary and the association distance increases without bound as the absolute temperature approaches zero.[5]

In previous work examining Brownian motion from an entropic perspective[6,7], we demonstrated the usefulness of assigning a relative entropy $S(r)$ to a pair of particles in a fluid separated by a scalar distance $r$, where particle A is regarded as fixed and B as free to assume any one of $\Omega$ positions on the surface of a sphere of radius $r$ surrounding B; $\Omega \propto 4\pi r^2$. $S(r)$, which according to the Boltzmann formula is $S(r) = k_B \ln(\Omega)$, and the relative energy $U(r)$, which is a potential of mean force, together determine the Helmholtz free-energy $W(r) = U(r) - TS(r)$, where $T$ is the absolute temperature and $k_B$ the Boltzmann constant. Here we regard the presence of a relative minimum in $W(r)$, which occurs over a temperature range $0 \leq T < T_c$, as a necessary condition for the existence of AB, with the temperature-dependent width of the free-energy well determining the region of phase space associated with AB. When A and B are a pair of spherical, oppositely charged ions in a dilute (e.g., aqueous) solution, the width of the well leads to the expression for the association distance of an ion pair, $q^2/(2k_BTD)$, first obtained by Bjerrum[4], where $q$ is the absolute value of the ionic charge and $D$ the dielectric constant of the solution.

The concept of two-particle association can be readily extended to establishing a criterion for the onset of phase condensation in a one-component fluid such as a rare gas or molten alkali halide and deriving a critical temperature or freezing temperature, respectively. Because the liquid-to-solid transition, in particular, is still not well understood from a statistical-thermodynamic perspective[8], a simple model for the freezing process would be desirable. In our model a phase transition occurs when an associated particle pair AB is created through the combined occurrence of (1) a horizontal inflection point at $R_c$ in a plot of $W(r)$ versus $r$ and (2) a fluid density resulting in an average separation of $R_c$ between any pair of nearest-neighbor particles. Application of this criterion to a molten alkali halide results in an expression for the freezing temperature, $T_F = q^2/(2k_BDs)$, similar to the melting temperature obtained by Kosterlitz and Thouless[9] for a two-dimensional plasma. Here $s$ is the sum of the crystal radii of the cation and anion for a given salt.

## Theoretical Results

*Helmholtz Free Energy*

Consider a fluid with molecule A fixed and molecule B located a scalar distance $r$ from A. The configurational partition function for this system is $Q(r) = \Omega(r)\exp[-\beta U(r)]$, where $\beta = 1/k_BT$ and $\Omega(r)$ is the degeneracy factor introduced earlier in conjunction with $S(r)$. The resulting Helmholtz free energy is, therefore, given by $W(r) = -\beta^{-1}\ln Q(r)$. The motivation for assigning an entropy $S$ to the pair of particles A,B is that their average scalar separation, $<r>$, gradually increases with time as a result of random thermal motion in a fluid (heat) bath of infinite volume. One may view this process as occurring because the number of possible configurational states increases with increasing $r$. Here the roles played by $r$ in determining the entropy and the potential of mean force are similar; in order to calculate the change in either quantity with $r$, $r$ must be varied reversibly (extremely slowly) so as to permit the statistical averaging of molecules in the surrounding medium. For this reason $W$ should be regarded as a free energy *of mean force*.

We illustrate the use of $W(r)$ first with the Lennard-Jones (LJ) potential applied to a noble gas because $W$ in this case can be solved analytically for its extremal points, thereby permitting the calculation of the equilibrium constant for a typical dimerization reaction, Ar + Ar ⇔ Ar$_2$, and providing an estimate for the critical temperature of an LJ-type gas. Thus, $U_{LJ}(r) = 4\varepsilon[(\sigma/r)^{12} - (\sigma/r)^6]$, and

$$W(r) = U_{LJ}(r) - 2k_B T \ln(r/\sigma). \tag{1}$$

The factor $1/\sigma$ in the logarithmic expression was chosen arbitrarily to render the independent variable dimensionless; $\sigma$ is a measure of the approximate diameter of the atom. Such an arbitrary constant does not affect the $W$-versus-$r$ behavior as far as the location of the extremal points is concerned; it merely shifts the curve vertically.

Figure 1 shows plots of $W$ versus $r/\sigma$ at three different temperatures, 0 K, $T_c/2$, and $T_c$, where $T_c$ is the critical temperature. Values of $1.5\varepsilon/k_B$ and $2^{1/3}\sigma$ were obtained for the critical temperature and critical separation, $R_c$, respectively, by solving

$$\left(\frac{\partial W}{\partial r}\right)_{T=T_c} = \left(\frac{\partial^2 W}{\partial r^2}\right)_{T=T_c} = 0. \tag{2}$$

It should be emphasized that $T_c$ is calculated for two atoms interacting in a vacuum, and to obtain a value appropriate for a dense gas at the critical point, $U_{LJ}(r)$ must be replaced by a potential of mean force calculated for two atoms interacting in an appropriate medium; see Appendix A.

*Equilibrium Constant*

The expectation value of the equilibrium constant for the reaction A + B ⇔ AB can be calculated from an expression of the type[1,2]

$$K_{AB} = 4\pi \int_{R_0}^{R_{max}} r^2 \exp[-\beta U(r)]dr, \tag{3}$$

valid for a dilute solution or gas, where the interval $R_0 \leq r \leq R_{max}$ defines the phase space associated with the dimer AB. Rather than taking $R_{max}$ to be an arbitrary cutoff distance in the effective range of the potential as is frequently done[1,2], we regard $R_{max}$ as the location of the relative maximum in $W(r)$, see curve (b) in Fig.1.

$$R_{max} = R_c[1 - (1 - T/T_c)^{1/2}]^{-1/6}. \tag{4}$$

Because of the steep repulsive potential present for $r < \sigma$, AB effectively resides in the region near $R_{min}$, the relative minimum in a free-energy well of width $h(T) = R_{max} - R_0$, where $R_0$ is a root of the equation $W(r,T) - W(R_{max},T) = 0$. The range for $R_{min}$, $2^{1/6} \leq R_{min}/\sigma < 2^{1/3}$, corresponds to $0 \leq T < T_c$. Curve (c) in Fig.1 indicates that $h(T_c) = 0$; at the critical temperature both $R_{max}$ and $R_0$ converge to $R_c$ and AB ceases to exist ($K_{AB} = 0$) in the thermodynamic sense. On the other hand, $R_{max} \to \infty$ as $T \to 0$. Eq. 3 could be used to calculate the equilibrium constant for the dimerization reaction of argon (2Ar ⇔ Ar$_2$) at low densities for $T < T_c$; $K_{Ar_2} = [Ar_2]/[Ar]^2$.

When A and B represent the pair of ions in a dilute solution described earlier, their interaction may be characterized by the so-called primitive model[5] where a pair of ions of equal and opposite charge are treated as hard spheres which interact with one another at long range by means of Coulomb's law. The ions are immersed in a dielectric medium whose static dielectric constant, $D$, is assumed to be the macroscopic value. The resulting potential may be written

$$U(r) = - q^2/Dr \qquad r \geq s,$$
$$U(r) = \infty \qquad r < s, \qquad (5)$$

where $s$ is the sum of the hard-sphere radii (closest approach). With this potential, $R_0 = s = R_c$; $R_{max} = \beta q^2/2D$; $T_c = q^2/(2k_B Ds)$ and the resulting width of the free-energy well becomes

$$h(T) = \frac{q^2}{2Dk_B}\left(\frac{1}{T} - \frac{1}{T_c}\right). \qquad (6)$$

$R_{max}$ is the distance that Bjerrum[1,4,5] designated as the separation beyond which an ion pair is no longer associated and can be regarded as a pair of free ions. Whereas Bjerrum selected this cutoff separation somewhat arbitrarily, in the present context it arises as the upper boundary of a relative minimum in the two-particle free energy; $R_{max} = h(T) + s$. Thus for a dilute solution of sodium chloride in water, the equilibrium constant for the reaction $Na^+ + Cl^- \Leftrightarrow NaCl_{aqueous}$ would be given by Eq. 3.

*Liquid-to-solid Transition*

Consider a molten salt such as an alkali halide or alkaline-earth oxide at a temperature slightly above its melting point. We shall assume that the potential given by Eq. 5 applies to ion pairs in a medium consisting solely of mobile ions. A static dielectric constant in Eq. 5 [rather than the optical value $D_\infty$, which is the square of the fused-salt refractive index at optical frequencies] is used because $U(r)$, as a potential of mean force, is the change in potential energy resulting when the separation of a pair of ions is reduced from infinity to $r$ in a reversible fashion [extremely slowly to maintain thermodynamic equilibrium]. Thus, the spatial location of the "medium" ions as well as their orbital-electron polarizabilities contributes to the total polarization. This positional contribution is analogous to the orientational polarization of dipoles in a "Debye" dielectric; hence $D > D_\infty$.

Just below $T_c$, a relative minimum in the free energy occurs for a pair of oppositely charged ions, and because the average nearest-neighbor separation for any pair of ions will be $\approx s$, if a pair of adjacent ions momentarily separates due to a thermal fluctuation, each ion will immediately be in the presence of a new ion of opposite charge. In other words, because of their proximity to one another, oppositely charged ions become trapped in the free-energy well, as it is now impossible to escape beyond $R_{max}$. Thus at $T_c$, the fluid becomes unstable with respect to the formation of bound ion pairs, and the liquid condenses to a solid. $T_c$ can be identified with the freezing temperature,

$$T_F = q^2/(2k_B D s). \tag{7}$$

Another way of viewing this transition is by noting that a force may be associated with the entropic term in Eq. 1; $f = T(\partial S/\partial r)_T = 2/\beta r$. If an ideal-gas particle were attached to the origin by means of an *imaginary* thread of length $r$, this force would be apparent as a tension in the thread of average magnitude $f$. At $T_F$, this repulsive force just equals the attractive coulombic force acting between the ion pair, resulting in Eq. 7; this equality of forces is, of course, a consequence of the first term in Eq. 2. Kosterlitz and Thouless[9] derived an expression very similar to Eq. 7 when describing the temperature at which ion pairs in a two-dimensional plasma separate, leading to a disordered system. Unfortunately, experimentally measured static dielectric constants for fused alkali halides and alkaline-earth oxides are not available in the literature, preventing a direct comparison between $T_F$ calculated from Eq. 7 and the observed melting temperatures. Appendix B discusses a means of estimating $D$ from heat-of-fusion data, thereby permitting an indirect comparison of theory with experiment.

## Conclusion

The view of molecular association presented in this article is predicated on an $r$-dependent, two-particle free energy, where $r$ plays the same role as it does in the usual definition of a potential of mean force. The presence of a relative minimum in $W(r)$, which by definition occurs for $T < T_c$, is a necessary condition for the existence of the adduct AB. $R_{max}$ is the distance such that if a single A,B pair were confined to a volume sufficiently small to ensure $r < R_{max}$, A and B would combine to form an AB molecule whose approximate size (*bond length*) would be $R_{min}$, the location of the minimum in the free-energy well. $R_{max}$ can also be viewed as the separation at which the $2/\beta r$ entropic repulsive force introduced earlier is balanced by the opposing mean force derived from $U(r)$. With the interval $R_0 \leq r \leq R_{max}$ defining the region of phase space associated with AB, the equilibrium constant can be calculated from Eq. 3 at low gas densities or solute concentrations for $0 << T < T_c$.

In general, the occurrence of chemical equilibrium or phase condensation will depend on the number density of the relevant species. Phase condensation will take place in a single-component fluid at $T < T_c$ if the average nearest-neighbor separation approaches $R_c$. On the other hand, chemical equilibrium will occur at low number densities in solution or in the gas phase. The use of an LJ potential with argon is heuristic; gaseous equilibria such as $2F \Leftrightarrow F_2$ or $2H \Leftrightarrow H_2$ characterized by a Morse potential would serve equally well.

We have suggested that Bjerrum's remarkable insight regarding the *association distance* ($R_{max}$) has a thermodynamic origin based on $W(r)$ and is relevant in describing phase condensation as well as chemical equilibrium. In our view, however, this distance is the radius of a *sphere of influence* with particle A at the center, rather than the actual size of the adduct, a distinction that is quite important at low temperatures. Finally, we have presented experimental evidence of an indirect nature for the liquid-to-solid transition as described by Eq. 7, in addition to pointing out the relevance of this equation to the result of Kosterlitz and Thouless.

**Appendix A:** *Observed properties of argon at the critical point.*

The observed values for the critical temperature and critical volume of argon are $1.3\varepsilon/k_B$ (151 K) and 75 cm$^3$/mole, respectively.[8] The latter enables an estimate for the average separation distance between nearest neighbors of 5.0 Å, which can be compared with $R_c = 4.6$ Å, obtained from Eq. 2 using $\sigma_{Ar} = 3.62$ Å[8]. The agreement between theory and experiment for both $T_c$ and $R_c$ is reasonable given the LJ potential was used in the calculation instead of a potential more appropriate for a dense gas.

In simple monatomic fluids with a pairwise additive intermolecular potential energy, the pair distribution function, $g(r)$, is related to the potential of mean force by $U(r) = - [\ln g(r)]/\beta$.[10] With this understanding, the condition for obtaining $T_c$ and $R_c$ given by Eq. 2 can be rewritten as $\partial^2 G/\partial r^2 = \partial G/\partial r = 0$, where the radial distribution function $G$ is defined by $G(r) = 4\pi r^2 g(r)$. The $G$-versus-$r$ behavior for argon in both the liquid and vapor phases was determined at 149 K, near the critical point, by x-ray diffraction measurements[11]; both curves show a horizontal inflection point in the region between 4 and 5 Å. The equation $\{W(r) = - [\ln G(r)]/\beta + \text{constant}\}$ expresses the fact that the free-energy (of mean force) bears the same relationship to $G(r)$ that $U(r)$ bears to $g(r)$.

**Appendix B:** *Estimation of the molten-salt static dielectric constant.*

Assuming the Coulombic interaction given by Eq. 5, the reversible dissolution of an ionic crystalline alkali halide or alkaline-earth oxide into a large volume of pre-existing melt at the melting temperature is analogous to the sublimation of the crystalline substance into a vacuum where ($D = 1$). In the latter case, the energy required is the crystal-lattice energy $\Delta H_L$, given by

$$\Delta H_L = \frac{1}{2} \sum_{i \neq j} q_i q_j / r_{ij}, \tag{B-1}$$

where $r_{ij}$ is the distance between the i$^{th}$ and j$^{th}$ ions as they appear in the crystal lattice. In the fusion process, the ions present in the crystal lattice diffuse into a medium of dielectric strength $D$; this dissolution is accompanied by the absorption of an amount of energy equal to the heat of fusion $\Delta H_F$. Thus,

$$\Delta H_F = \frac{1}{2} \sum_{i \neq j} q_i q_j / D r_{ij} = \Delta H_L / D, \qquad (B-2)$$

and $D = \Delta H_L / \Delta H_F$. Representative values of $D$ are 28 and 52 for KCl and MgO, respectively, as determined from their lattice energies and heats of fusion[12,13]. The observed melting temperatures[12,13] and those calculated from Eq. 7 are 1043 and 1023 K, respectively, for KCl; and 3173 and 3142 K, respectively, for MgO.

**Figure Caption**

*Figure 1.* The two-particle free energy in units of $\varepsilon$ is depicted as a function of the reduced separation distance $r/\sigma$ for three different temperatures, based on Eq. 1: Curve (a), 0 K; Curve (b), $T_c/2$; and Curve (c), $T_c$.

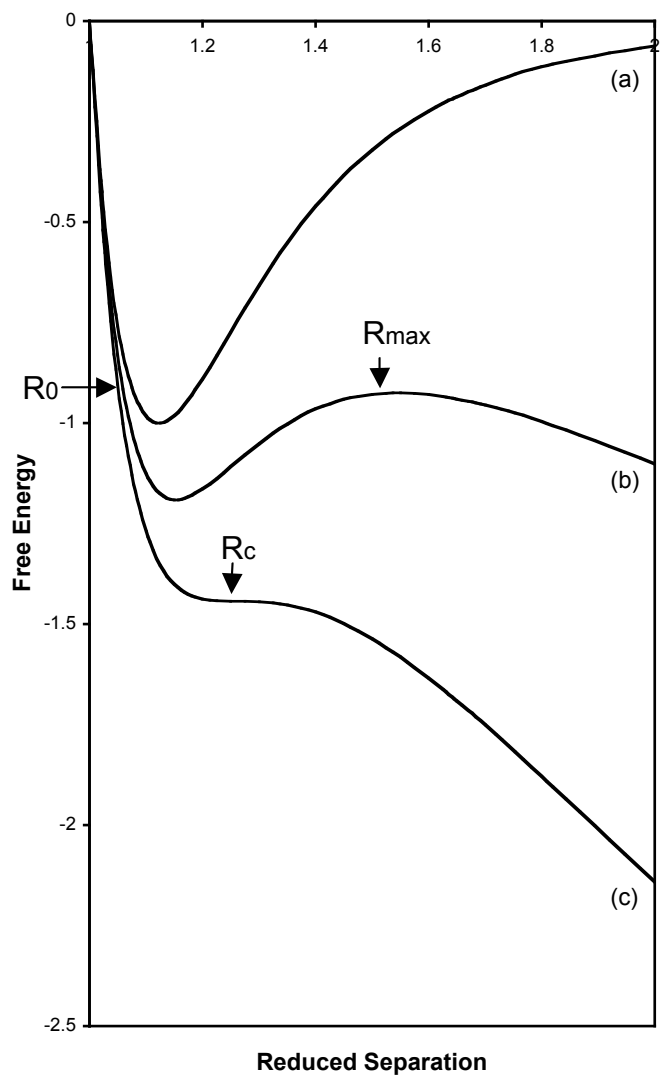

Figure 1.